\documentclass[]{emulateapj}
\pdfoutput=1

\usepackage{graphicx}
\usepackage{natbib}
\def\beq{\begin{equation}}
\def\eeq{\end{equation}}
\long\def\symbolfootnote[#1]#2{\begingroup%
\def\thefootnote{\fnsymbol{footnote}}\footnote[#1]{#2}\endgroup}

\begin{document}
\title{Induced Rotation in 3D Simulations of Core Collapse Supernovae: Implications for Pulsar Spins}

 \author{Emmanouela Rantsiou\altaffilmark{1},
   Adam Burrows\altaffilmark{1},
   Jason Nordhaus\altaffilmark{1},
   Ann Almgren\altaffilmark{2}}

\altaffiltext{1}{Department of Astrophysical Sciences, Princeton University, Princeton, NJ 08544 USA;
emmarant@astro.princeton.edu, burrows@astro.princeton.edu, nordhaus@astro.princeton.edu}
\altaffiltext{2}{Computational Research Division, Lawrence Berkeley
National Lab, Berkeley, CA 94720, USA; asalmgren@lbl.gov}

\begin{abstract}
It has been suggested that the observed rotation periods of
radio pulsars might be induced by a non-axisymmetric spiral-mode
instability in the turbulent region behind the stalled supernova bounce shock,
even if the progenitor core was not initially rotating. In this paper,
using the three-dimensional AMR code CASTRO with a realistic progenitor
and equation of state and a simple neutrino heating and cooling scheme,
we present a numerical study of the evolution in 3D of the rotational profile
of a supernova core from collapse, through bounce and shock stagnation, to
delayed explosion. By the end of our simulation ($\sim$420 ms after
core bounce), we do not witness significant spin up of the proto-neutron
star core left behind. However, we do see the development before
explosion of strong differential rotation in the turbulent gain
region between the core and stalled shock. Shells in this region acquire
high spin rates that reach $\sim$$150\,$ Hz, but this region contains
too little mass and angular momentum to translate, even if left behind,
into rapid rotation for the full neutron star. We find also that much
of the induced angular momentum is likely to be ejected in the
explosion, and moreover that even if the optimal amount of induced
angular momentum is retained in the core, the resulting spin period
is likely to be quite modest. Nevertheless, induced periods of seconds
are possible.
\end{abstract}

\section{Introduction}

One of the key questions associated with radio pulsars
is the origin of their spin periods (Lorimer 2009,2010). In two recent papers
(Blondin \& Mezzacappa 2007 [hereafter BM07]; Blondin \& Shaw 2007),
it has been suggested that the growth in 3D of an $\ell = 1$ non-axisymmetric
($m = \{-1,1\}$) spiral mode (Fern\`andez 2010) of the so-called standing accretion shock
instability (``SASI") during the post-bounce delay phase seen in current
supernova theory can spin up matter in the turbulent region behind the stalled
shock.  BM07 and Blondin \& Shaw (2007) find that the accretion of this spinning 
matter (and its associated angular momentum) onto the inner core can spin up
the nascent neutron star to rotation periods that compare favorably
with the observed/inferred values for pulsars at birth.
More precisely, they report final periods of $\sim$70 milliseconds (ms)
for moderately rotating progenitors and $\sim$50 ms for non-rotating
progenitors. The measured period range of birth periods is from $\sim$10 
milliseconds to seconds, with an average ``injection period" near 
$\sim$0.5 seconds (Chevalier \& Emmering 1986;  Narayan 1987)\footnote{The inferred initial spin period
of the Crab pulsar is 15$-$16 milliseconds (Atoyan 1999), but the average for
radio pulsars is considerably longer.}. If this mechanism were to obtain,
it would obviate the need to invoke a spinning progenitor core.  However,
rotation is a natural feature of stars, and there is no obvious reason
the progenitor ``Chandrasekhar" mass in the center of a massive star at
death wouldn't be rotating with the spins and angular momenta needed to
leave a compact object with the measured spin frequencies (Ott et al. 2006ab).
This is a function of the angular momentum distribution in the cores
of massive stars when they collapse (Heger, Woosley, 
\& Spruit 2005; Maeder \& Meynet 2000,2004; Hirschi,
Meynet, \& Maeder 2004). Hence, the possibility that the 3D dynamics of the core
itself could generate countervailing flows with opposite signs of
angular momentum, ``one stream" of which could leave the residue spinning
at interesting rates while conserving overall angular momentum, is
intriguing.  If this were true, to zeroth order all cores at collapse
could be non-rotating, and the spins could be a function of collapse dynamics alone.
This would eliminate one of the current ambiguities in massive star progenitor
models, whose core spins at collapse are not observable.  The progenitor cores
could all be born with ``zero" angular momentum, while giving birth to
proto-neutron stars with respectable spins.

However, BM07 and Blondin \& Shaw (2007) employed various simplifications
in their simulations.  They did not follow collapse itself, but started
in a steady-state post-bounce configuration.  They used a simple ``gamma-law" equation
of state (EOS), and not a full nuclear EOS.  They set the mass accretion rate
through the shock to a constant value and did not perform their simulations in the context
of a realistic massive-star progenitor.  Importantly, they excised the inner
core whose very spin-up was being studied.  Rather, they inferred from
ostensible accretion of matter and angular momentum at their inner boundary
an irreversible accumulation of spin in the proto-neutron star.  Finally,
they did not consider the potential role of the supernova explosion itself
on the magnitude of the residual angular momentum and final induced core period.
They did, however, investigate the potential role of initial rotation in the
final outcome.

In this paper, we investigate the claims of BM07 and Blondin
\& Shaw (2007), improving upon their study in several ways.
Our 3D simulation is described in Nordhaus et al. (2010a) and
was carried out using CASTRO (Almgren et al. 2010), an Eulerian Adaptive Mesh
Refinement (AMR) hydrodynamics code with hierarchically nested
rectangular grids that are refined both in space and time.
We employed a realistic EOS (Shen et al. 1998ab) and a simple neutrino
transfer scheme (see Nordhaus et al. 2010a and Murphy \& Burrows 2008).
Both Nordhaus et al. (2010a) and Murphy \& Burrows (2008) incorporated 
the Liebend\"orfer (2005) prescription for electron capture on infall
and a simple neutrino heating and cooling algorithm after bounce.
The driving electron and anti-electron neutrino luminosities were
kept constant at $1.9\times 10^{52}$ ergs s$^{-1}$. This enabled us
to heat the so-called ``gain region" (Bethe \& Wilson 1985) and
investigate the pre-bounce infall, bounce, post-bounce delay, and
explosion phases. In this simulation (model 
3d:L\_1.9 of Nordhaus et al. 2010a), a neutrino-driven explosion
occurred approximately $\sim$200$-$250 ms after bounce and we carried the
simulation out to $\sim$420 ms after bounce.  Our initial model was the 15-M$_\odot$
solar-metallicity, non-rotating red-supergiant model of \cite{woos95}.
An important aspect of our simulation is that the proto-neutron star itself
is at all times present on the grid (which is Cartesian)
and is followed throughout the run in a hydrodynamically
consistent fashion.

The cubic grid in our simulations extended to $10^4\,$ kilometers (km) on each side,
with $304^3$ cubic cells covering the grid's volume at its coarsest
level. Three levels of refinement of a factor of $4$ in each dimension
were followed.  As a result, the resolution in the central
regions (the inner $200\,$km) was $\sim$0.5 km. The monopole approximation
for self-gravity was adopted.

By the end of our simulation, we find that there is almost no spin up of the proto-neutron
star. During the post-bounce phase and before explosion,
the region exterior to the core and interior to the shock ($\sim$60 to $\sim$250 km)
does witness the growth of spiral modes, some spin up (to $\sim$150 Hz), and the
emergence of countervailing rotational flows. However, the explosion ejects
not only mass, but its associated angular momentum, and depending upon
where the mass cut is finally imposed, the residual core can end up
with some net rotation. In our most optimistic scenario
the core achieves a period of only $1.2\,$seconds.  This fastest possible
spin does not alone seem adequate to explain the measured and/or
inferred pulsar birth periods without significant initial rotation
in the progenitor itself. In \S\ref{results}, we present
our quantitative results and we follow in \S\ref{discussion} with an extended
discussion and our summary conclusions.

\section{Results}
\label{results}

\subsection{Profiles and Temporal Evolution of Angular Momentum}
\label{profiles_L}

\begin{figure}
\begin{center}
%\ContinuedFloat
\includegraphics[height=0.4\columnwidth, width=0.75\columnwidth]{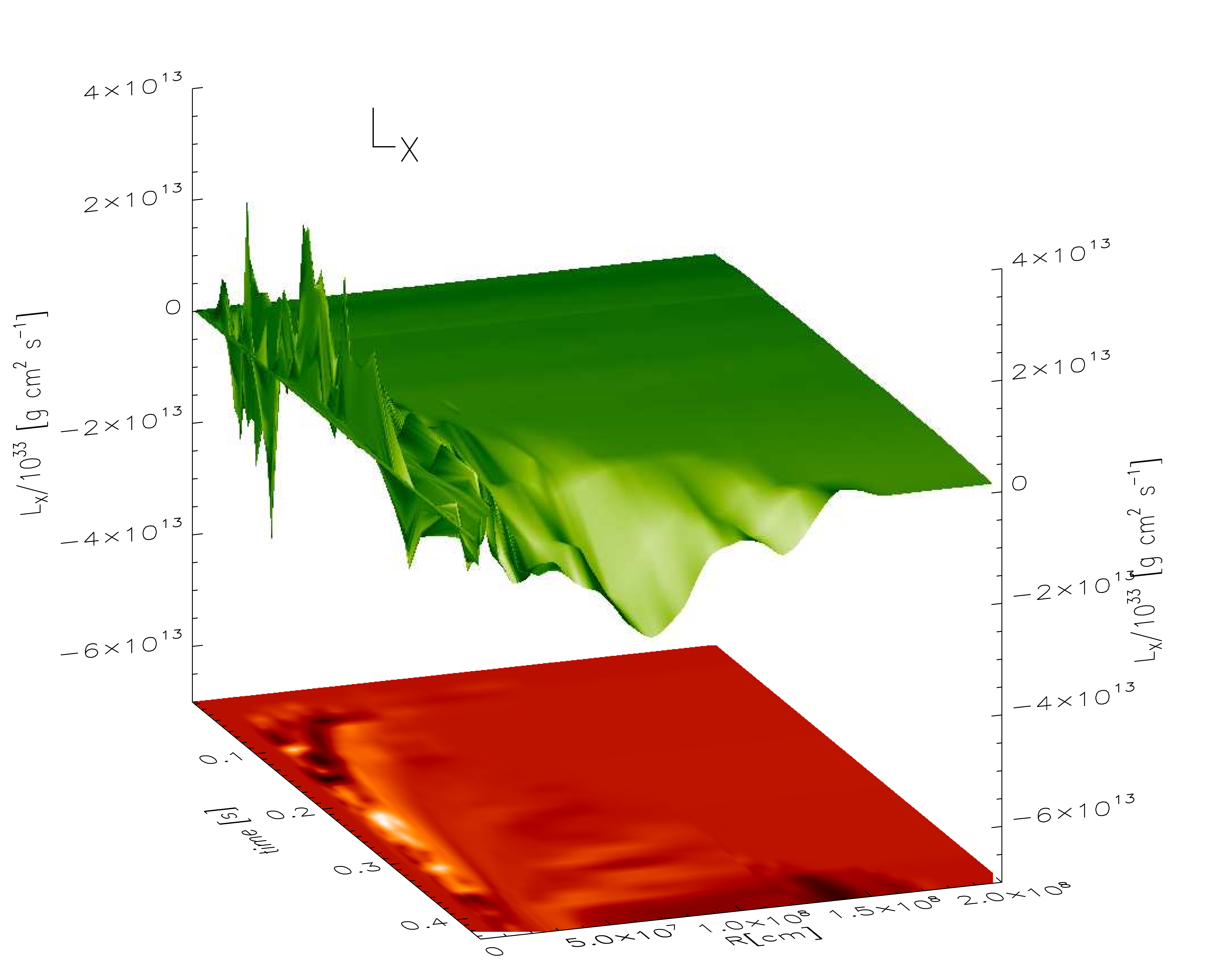}\\
\includegraphics[height=0.4\columnwidth, width=0.75\columnwidth]{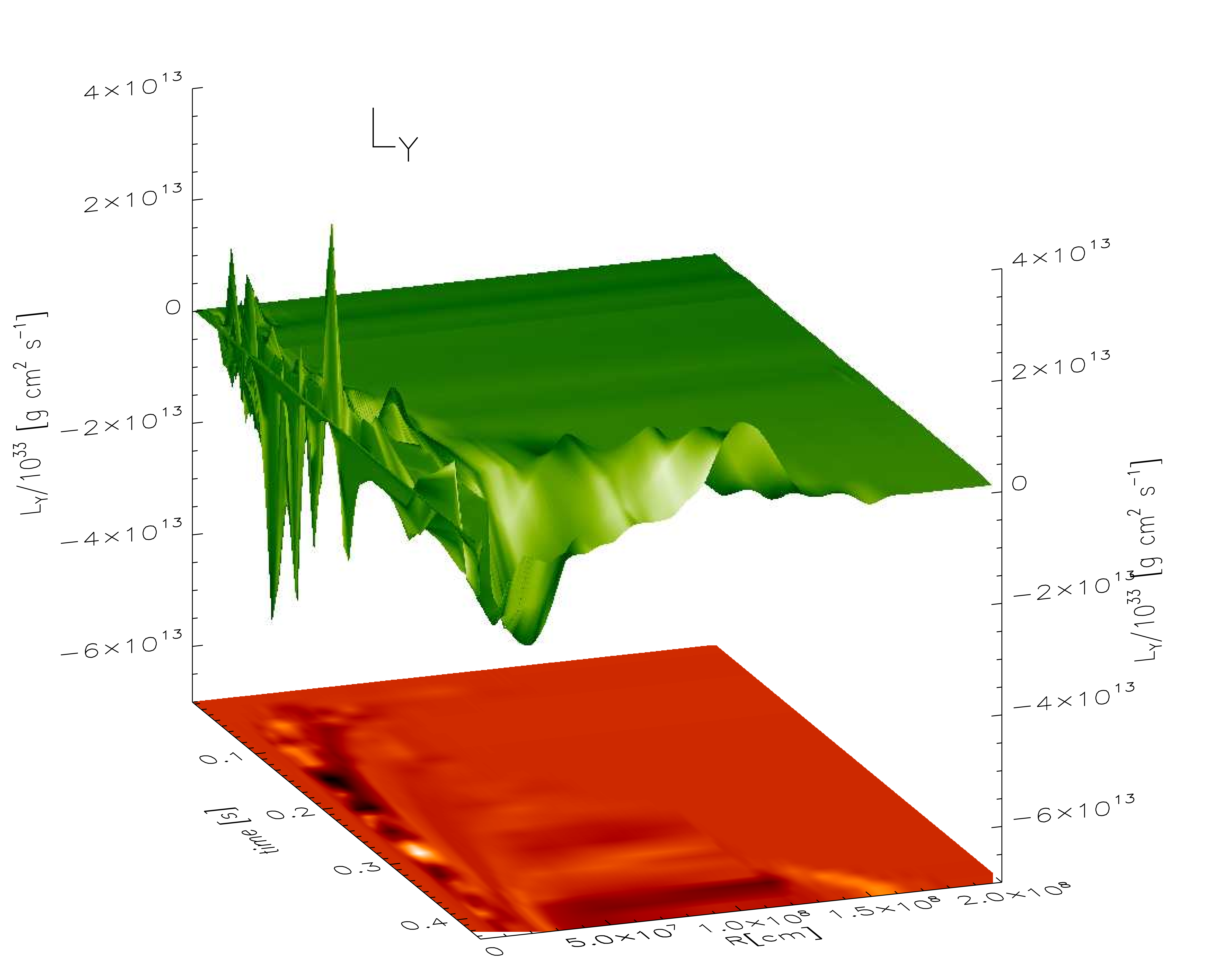}\\
\includegraphics[height=0.4\columnwidth,
width=0.75\columnwidth]{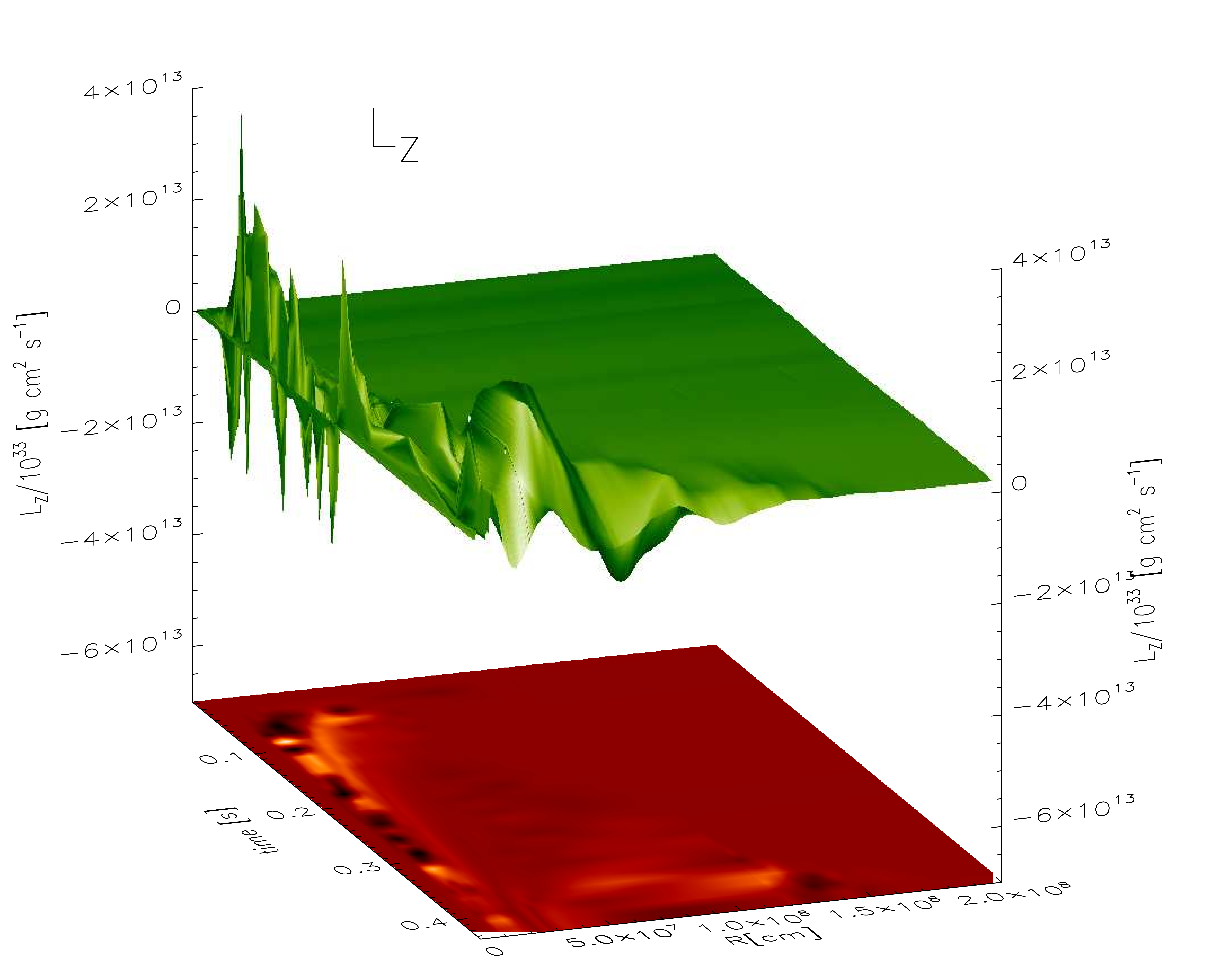}
\end{center}
\caption{Three-dimensional surface representation of the 
  temporal evolution (after bounce) of the components ($L_i$) of the total angular 
  momentum (vertical axes) enclosed within a spherical radius R (in km). The three 
  panels, from top to bottom, correspond to the x-, y-, and z-components 
  of the enclosed angular momentum. The two-dimensional plot under
  each surface shows the projection of the upper surface onto
  the plane, with the color variation indicating the value of the
  projected point, going from black (negative) to white 
  (positive) through a scale of red shades. We notice the initial increase of the angular momentum
  components in the very inner regions of our computational domain, and the
  ejection (with the ejected mass) of the bulk of angular momentum outwards during the later
  stages of the simulation. See text in \S\ref{profiles_L} for further discussion.}
\label{plot1}
\end{figure}

\begin{figure}
\begin{center}
\includegraphics[height=0.4\columnwidth, width=0.75\columnwidth]{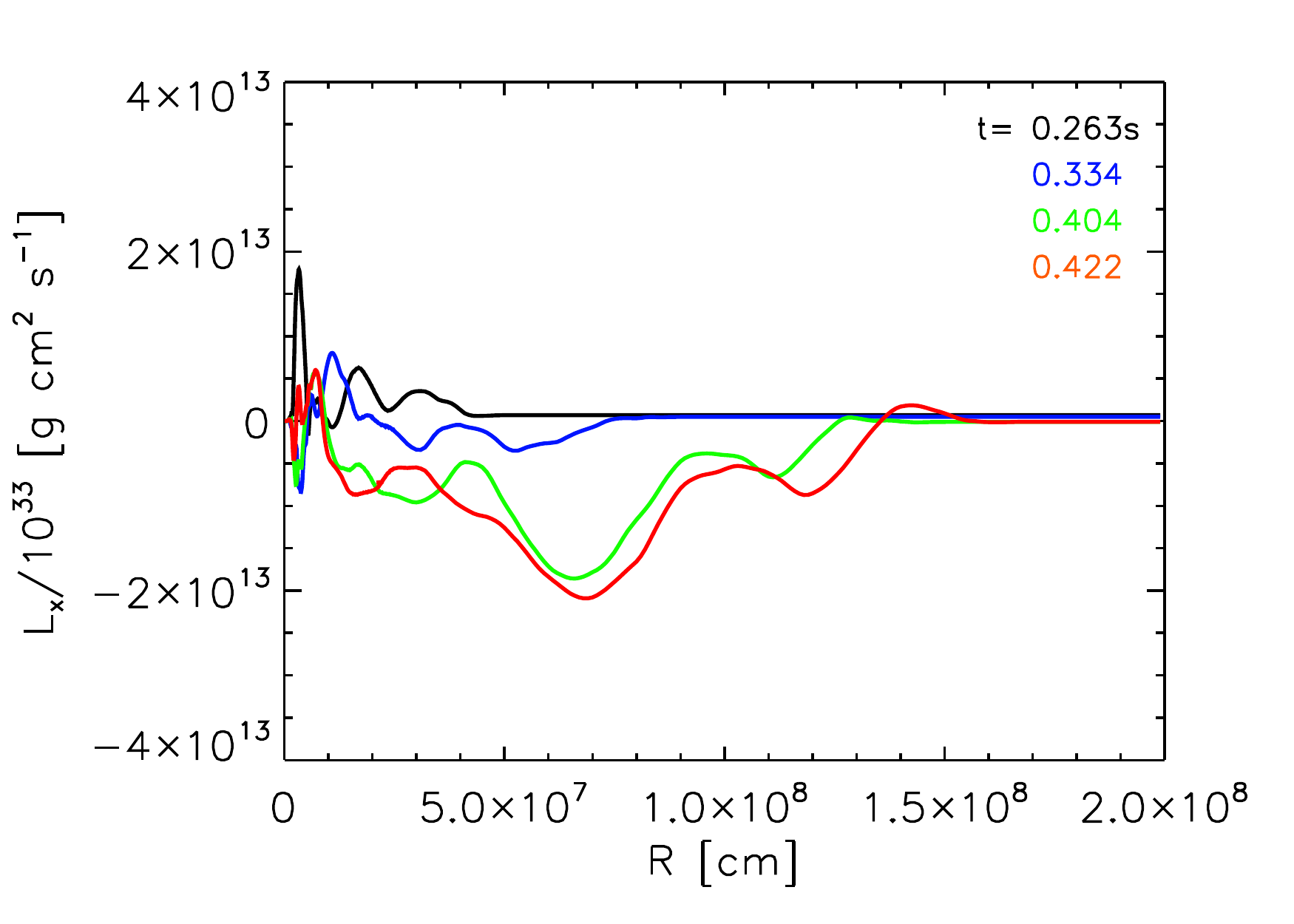} \\
\includegraphics[height=0.4\columnwidth, width=0.75\columnwidth]{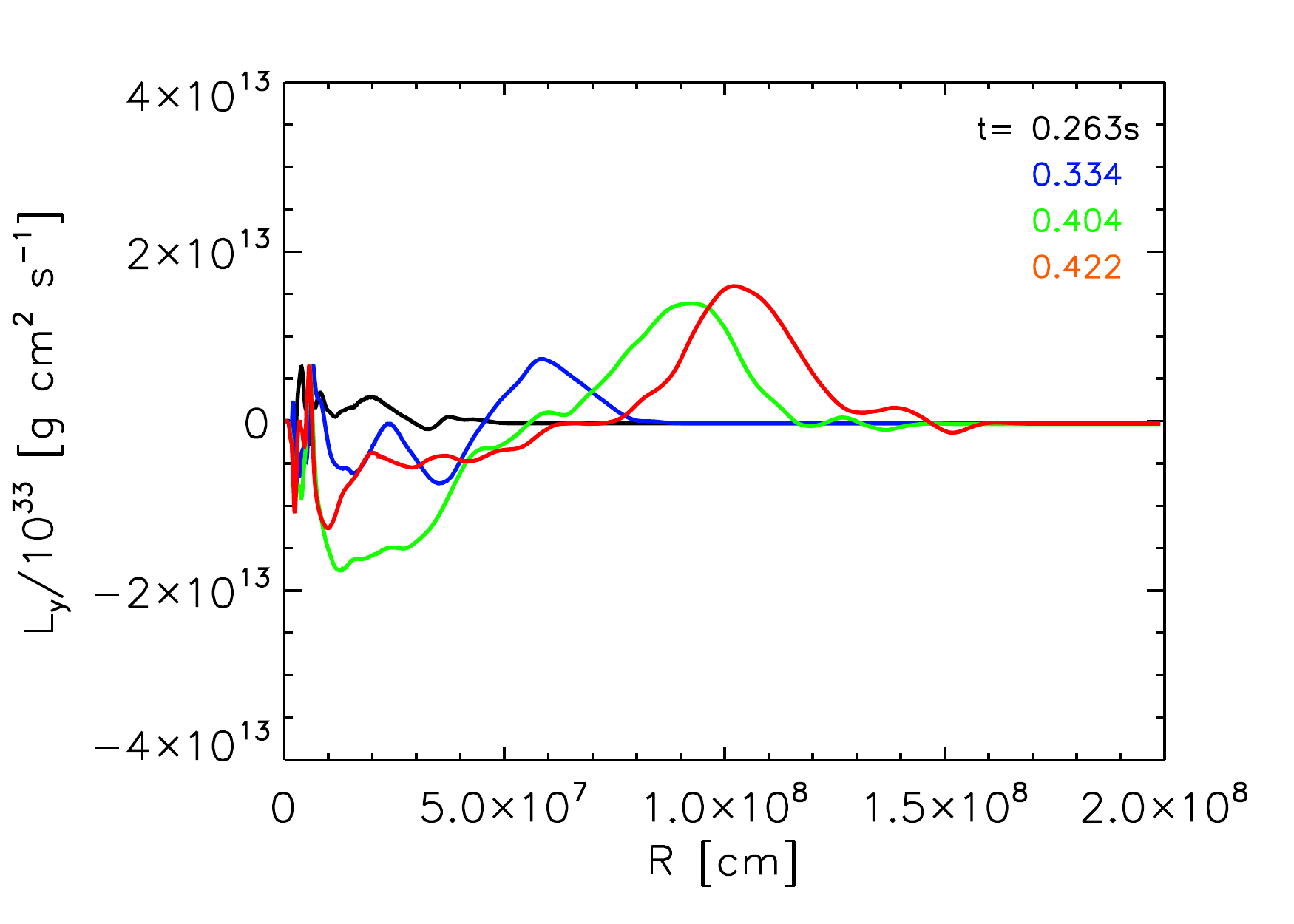} \\
\includegraphics[height=0.4\columnwidth,
width=0.75\columnwidth]{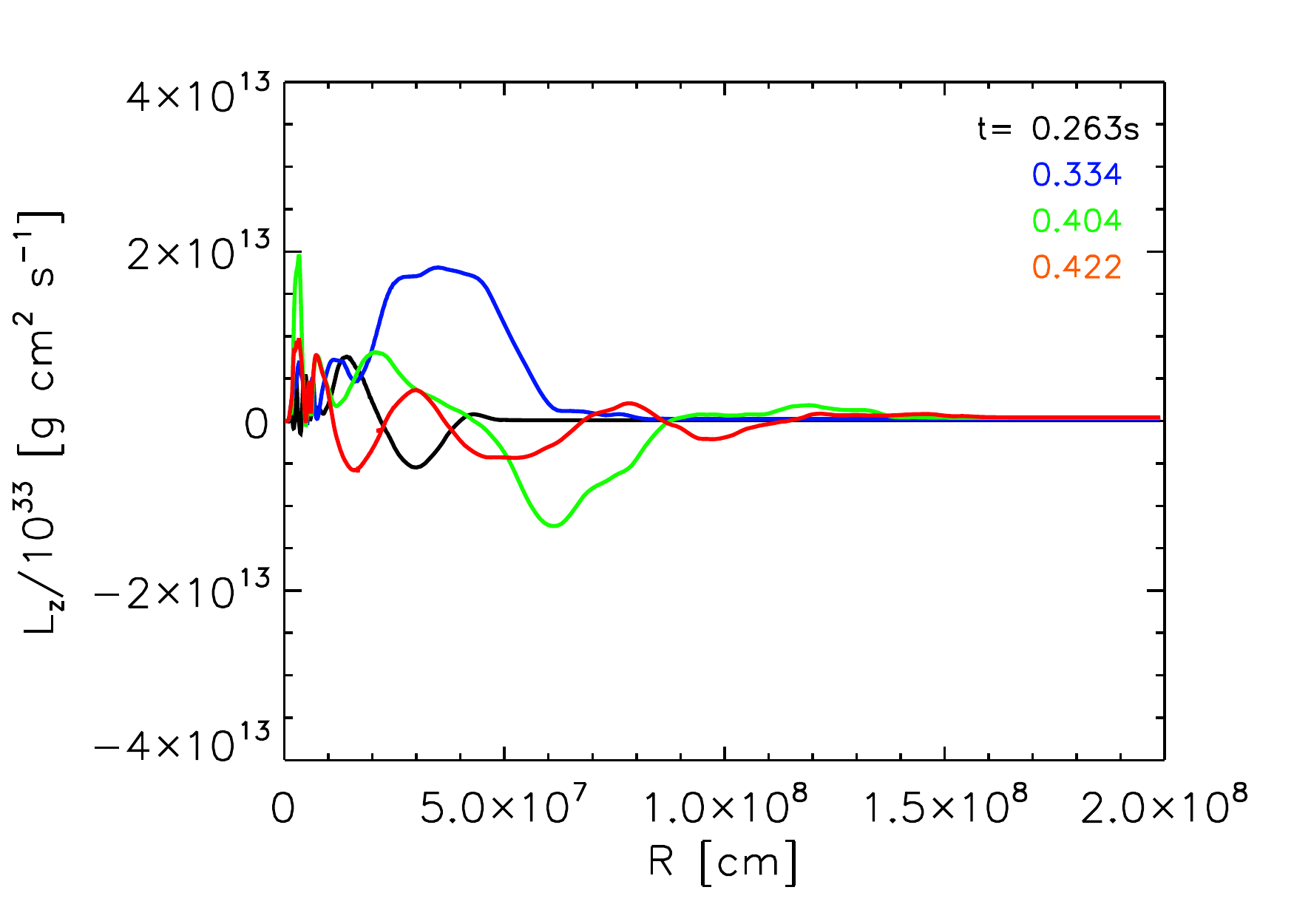}
\end{center}
\caption{Total angular momentum enclosed within a spherical radius R and its
  evolution with time after bounce, for four different timesteps. The three panels correspond to the three
  components of the angular momentum vector (x, y and z components
  from top to bottom, respectively). Lines of different color
  indicate different post-bounce times during the simulation. 
  We notice again (as in Fig.\ \ref{plot1}) an initial increase of the angular momentum 
  components between $\sim$60 and $\sim$200 km.
  With time, much of the angular momentum is ejected with the exploding mass. This can
  be seen here in the propagating angular momentum ``bump." 
  As discussed in the text and also in the captions to Table \ref{table1} and Fig.\ \ref{plot7},
  this ejected angular momentum, if actually all accreted onto the neutron star at 
  later time, would induce a rotational period no faster than $1.2\,$s.}
\label{plot2}
\end{figure}

We begin our analysis by calculating the amount of angular momentum
present in our computational domain. We divide the innermost
$2000\,$km of the grid into concentric spheres of increasing radius, and we
monitor the magnitude of each of the three components 
($x, y, {\rm and}\ z$) of the angular momentum ($L_i$) enclosed within each 
sphere for the duration of the calculation after core
bounce. The panels in Fig. \ref{plot1} depict three-dimensional 
surface representations of the temporal evolution of the three orthogonal 
components of the total angular momentum enclosed within a certain radius, as a 
function of that radius.  Such panels summarize what we find in our simulation. 
As Fig. \ref{plot1} indicates, we do indeed see an initial increase of angular 
momentum, but only between $\sim$60 km and  $\sim$250 km, and quasi-oscillatory behavior with a
period near $\sim$30 ms. The innermost zones that might be identified 
with the proto-neutron star are not spun up to periods faster than 
$\sim$5$-$10 seconds.  At later times (the last $0.17\,$ s after bounce), 
we note that the angular momentum that originally resided in this inner region is 
ejected with the exploding mass and follows the outward progress of the reenergized shock.  
This ejected angular momentum ``bump'' can also be seen clearly in Fig.\ \ref{plot2}, where
we show the spatial profile of all three components of angular
momentum for four different times near the end of our simulation. We address 
the subsequent fate of this ejected angular momentum in \S\ref{final_state} 
and \S\ref{discussion}. At the end of our simulation, very little 
angular momentum is left in the central region of our grid, indicating 
that the inner core is left rotating very slowly.  At the final timestep,
the spin periods in the inner ten kilometers are all greater than $\sim$10 seconds.
Note that the flatness of the regions in Fig. \ref{plot1} at large distances 
demonstrates that total angular momentum is very well conserved during the simulation.
In CASTRO, angular momentum is not conserved by construction, so that its
global conservation during this simulation is encouraging.  Specifically, before 
and after bounce the total angular momentum of matter on the grid is conserved 
to better than $\sim$1\%, staying very near ``zero" for the duration of the calculation.

\begin{figure}
\begin{center}
%\ContinuedFloat
\includegraphics[height=0.4\columnwidth, width=0.75\columnwidth]{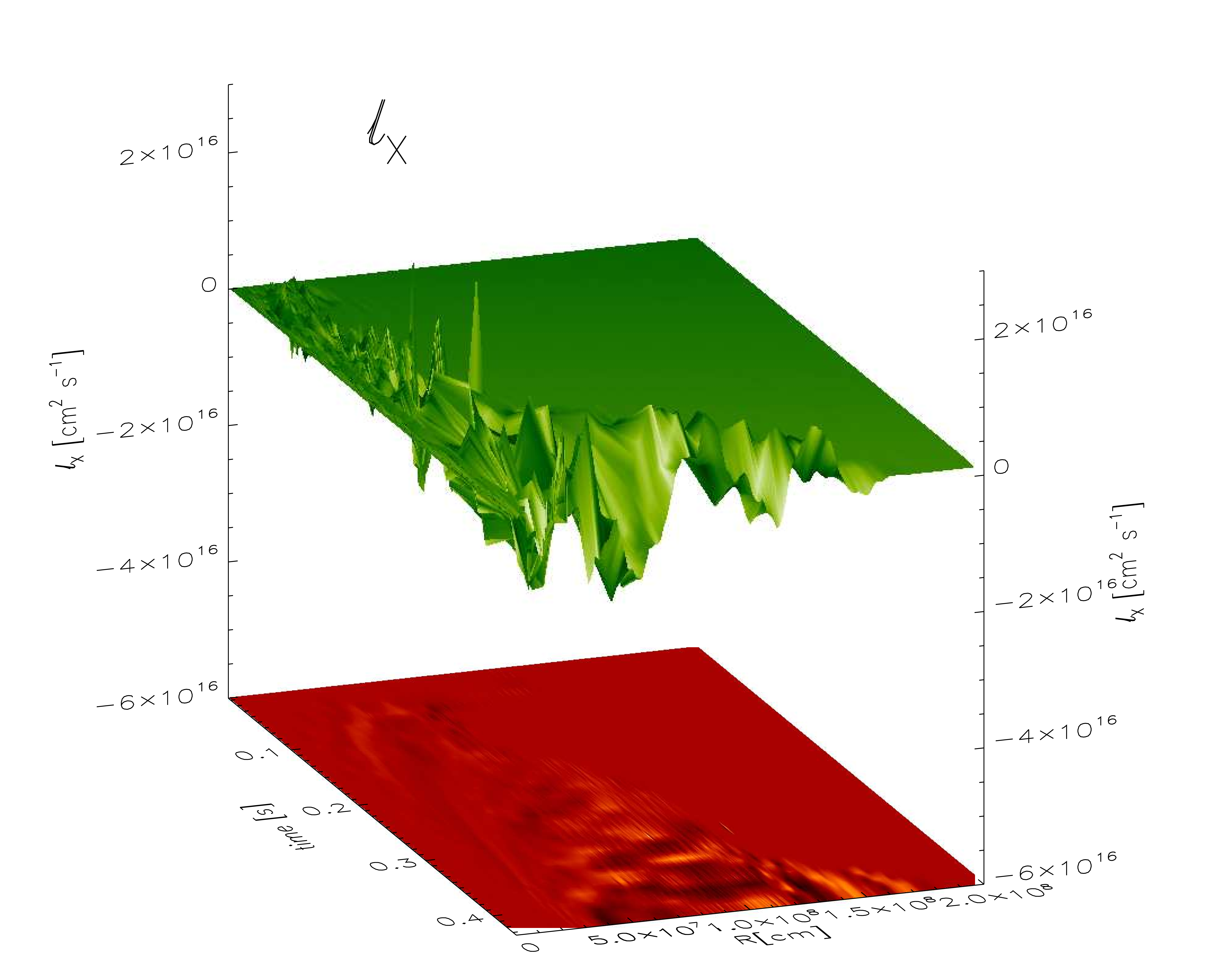}\\
\includegraphics[height=0.4\columnwidth, width=0.75\columnwidth]{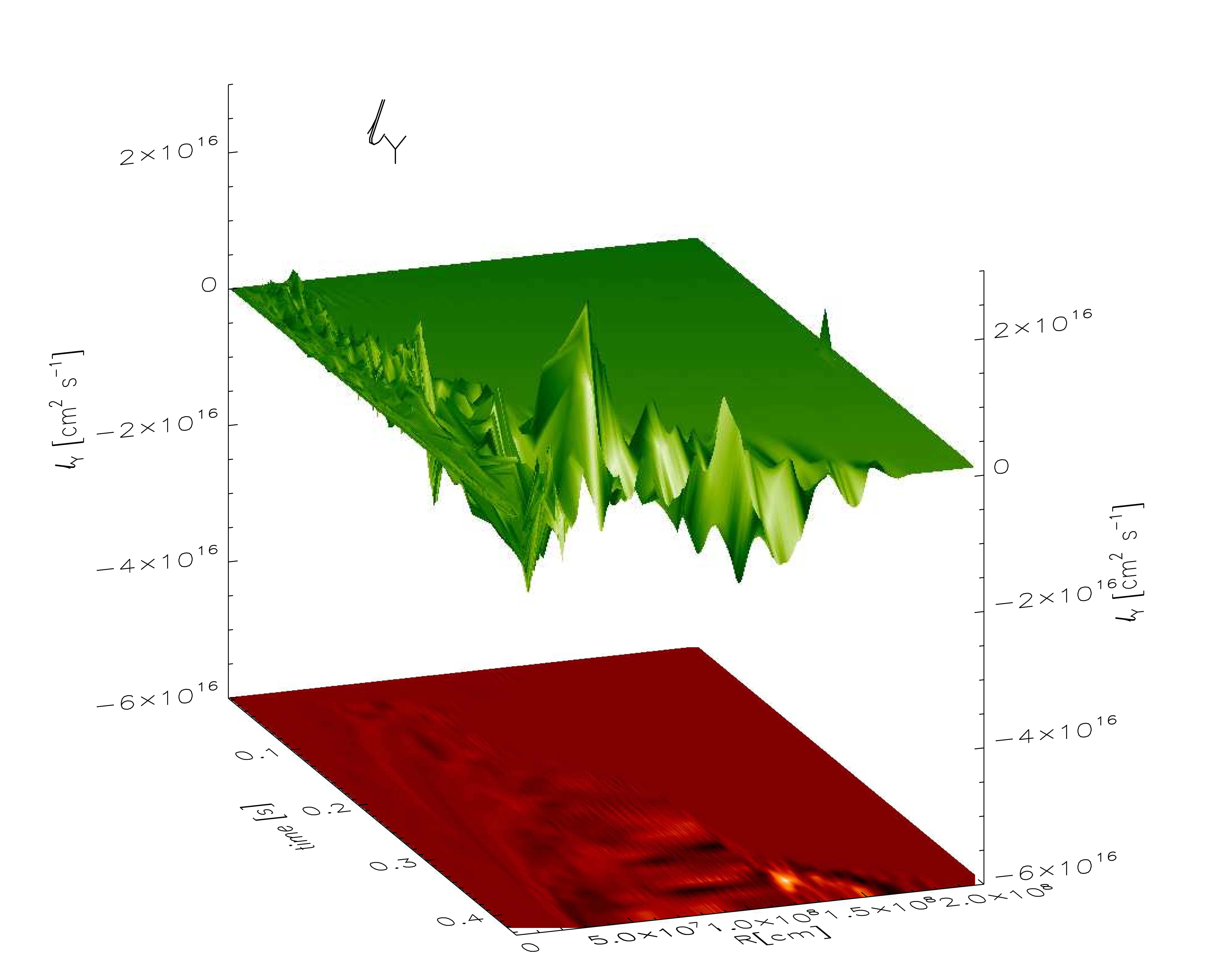}\\
\includegraphics[height=0.4\columnwidth,
width=0.75\columnwidth]{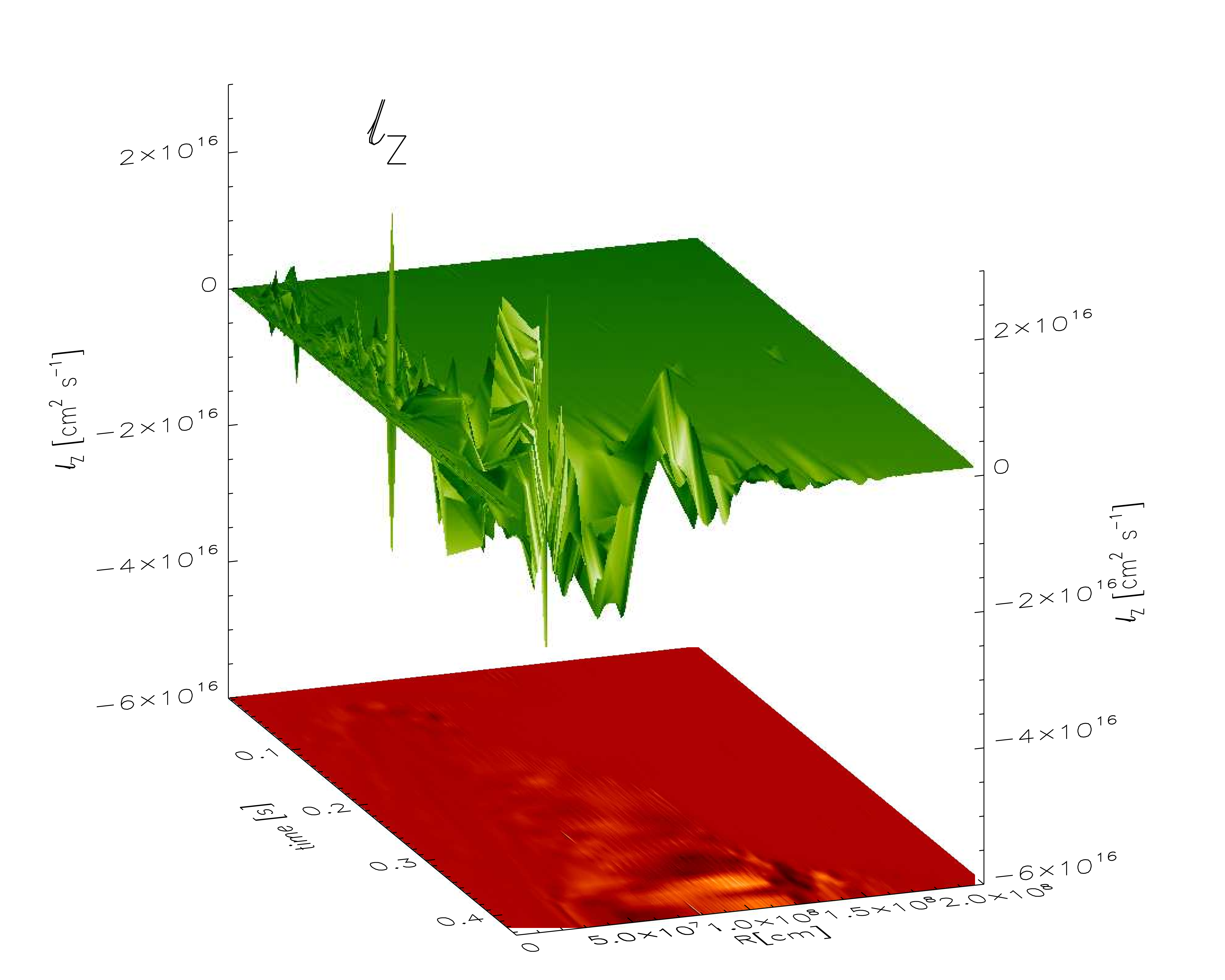}
\end{center}
\caption{Temporal evolution (in seconds after bounce) of the 
  average specific angular momentum ($\ell_i$, vertical axes) of
  spherical shells at various enclosed spherical radii (R, in centimeters). 
  R is logaritmically distributed. From top to bottom, the three
  panels correspond to the x-, y-, and z-component of the specific 
  angular momentum. To determine the averages, the computational grid 
  was divided into thin spherical shells of width $1\,$km within
  the inner $250\,$km of the grid and of width $10\,$km for radii between $250$ 
  and $2000\,$km. The two-dimensional surfaces under each 
  upper surface are projections of the upper surfaces, 
  with the color variation indicating the value of the
  projected point, going from black (negative) to white 
  (positive) through a scale of red shades. The propagation outward 
  at later times of the induced angular momentum is manifest in 
  these panels as well. See the text in \S\ref{profiles_L} for a discussion.}
\label{plot3}
\end{figure}

Next, we divide our grid into thin spherical shells that cover the innermost
$2000\,$km of the computational domain. For the inner
$250\,$km the width of the shells was chosen to be $1\,$km and for
radii between $250$  and $2000\,$km the width was increased to
$10\,$km. For each of the shells we calculated the average specific
angular momentum ($\ell_i$, not the total enclosed angular momentum), 
for all three components of the angular momentum.  Figure \ref{plot3} 
depicts their associated spatial and temporal profiles.  As discussed above and can be seen 
in Figs. \ref{plot1} and \ref{plot2}, regions interior to $\sim$200 km  
experience some growth of specific angular momentum during the 
earlier stages after core bounce, whereas the specific angular momentum 
at larger radii remains close to zero until after the onset of explosion.  
As Fig. \ref{plot3} makes clear, the sign of the specific angular momentum 
for each component varies with time and radius, conserving total angular momentum 
after bounce to high precision.  After the onset of explosion, 
angular momentum is seen to propagate outward with the ejected mass
and the shock wave.

An interesting feature is observed for radii between $60$ and $250\,$km. 
As time progresses in the simulation, the magnitudes of the specific 
angular momenta on shells in this region grow with time, indicating 
that in this region there is indeed some spin up by the spiral mode 
identified in BM07, Blondin \& Shaw (2007), and Fern\`andez (2010).  
We proceed in \S\ref{rot_freq} with a discussion of this phenomenon
and the associated spin periods.

\subsection{Rotational Period and Frequency Profiles}
\label{rot_freq}

%In Fig.\ \ref{plot4} we show the frequency profiles of solid spheres
%of various radii, for four different timesteps towards the end of the
%simulation. The timesteps are the same ones chosen at Fig.\
%\ref{plot2}. By knowing the enclosed angular momentum within each
%sphere (shown in Fig.\ \ref{plot2}) and assuming solid body rotation,
%we calculate their rotational frequencies. We notice that by the end
%of the simulation the magnitude of frequency for spheres of radii
%between $10 - 20\,$km does not exceed $\sim$0.32 Hz, giving a maximum
%rotational spin period of $\sim$3 s for a $20\,$km radius NS.

\begin{figure}
\begin{center}
%\ContinuedFloat
\centerline{\includegraphics[height=0.55\columnwidth, width=0.55\columnwidth]{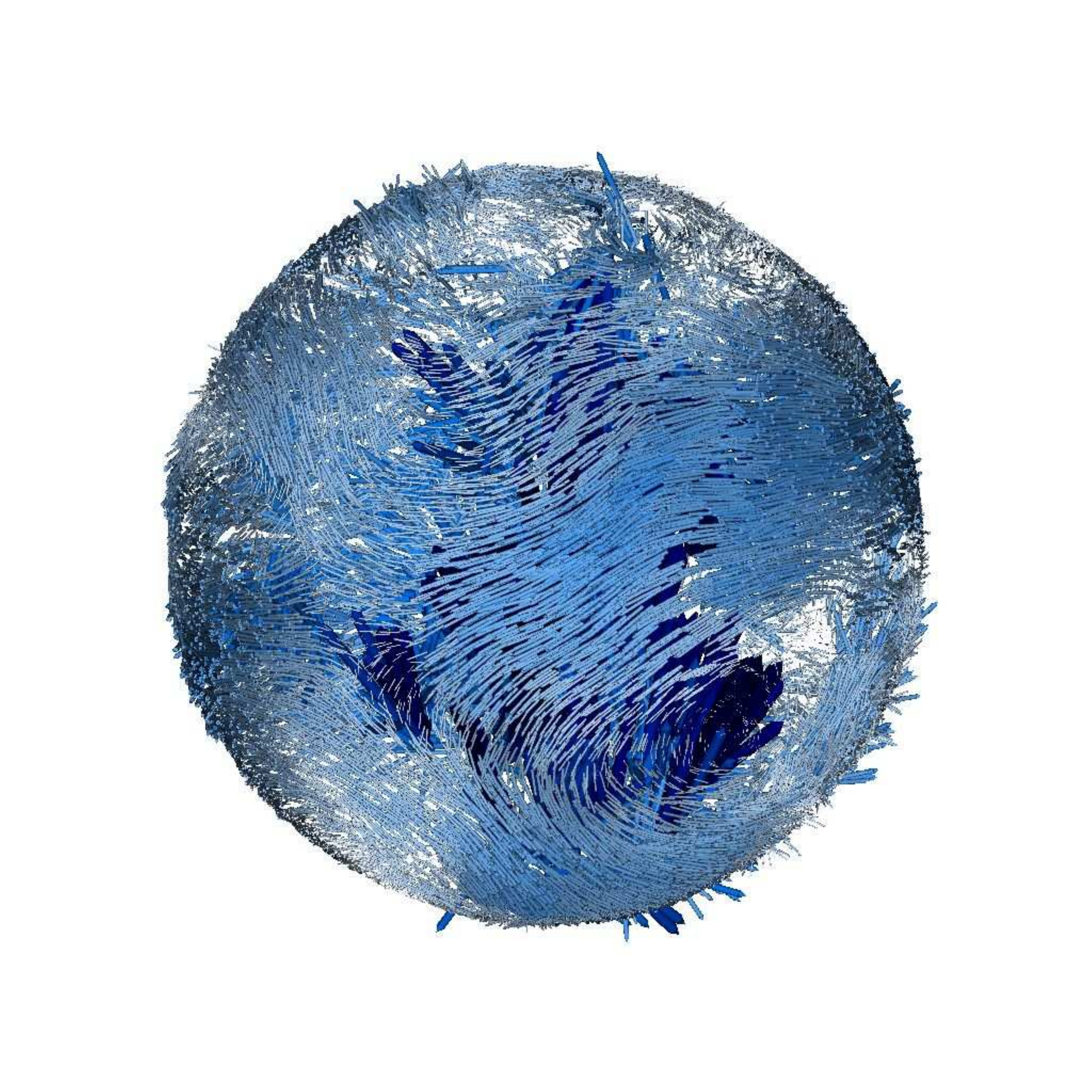}
\includegraphics[height=0.55\columnwidth, width=0.55\columnwidth]{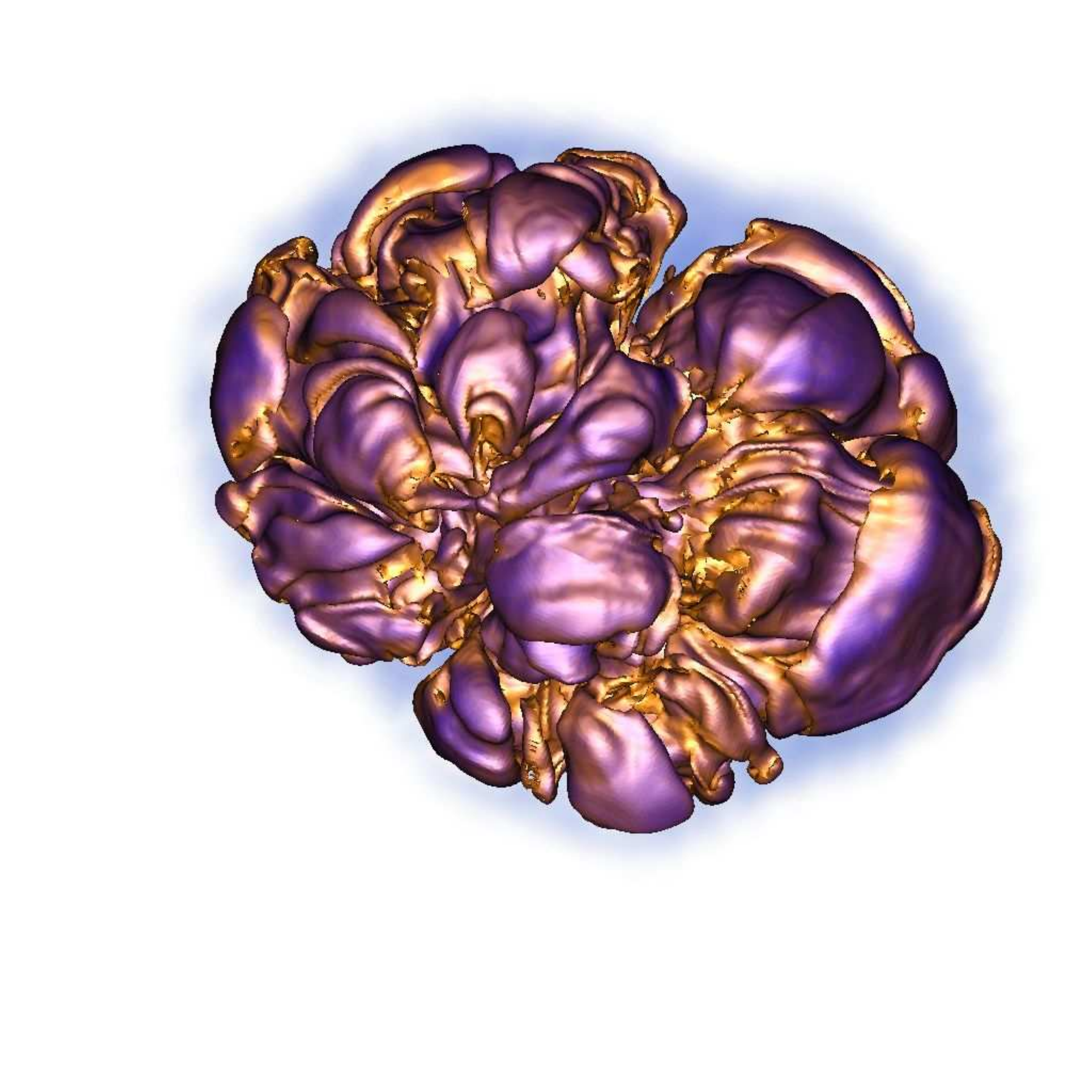}
}
\end{center}
\caption{{\bf Left:} Velocity vector field profile on a spherical shell of radius
  $90\,$ km towards the end of the simulation ($\sim$370 ms after core
  bounce). Both the size and color of the vectors represent the magnitude of
  the velocity field, with darker (lighter) shades corresponding to
  higher (lower) values. {\bf Right:} 3D rendering of an isentropic surface 
  with entropy equal to 10 $k_b$ baryon$^{-1}$. The colormap represents the
  magnitude of the entropy gradient on the surface, with values
  increasing going from orange to purple. The surface spans 1600 km 
  and corresponds to a time 380 ms after core bounce.}
\label{plot5}
\end{figure}

Having calculated the angular momentum contained in spherical shells
at various radii, we proceed to extract the rotational profiles of
those shells.  Figure \ref{plot4} shows the temporal evolution of the
average spin frequency ($f_i$, in Hz) of each shell, for rotation around the x, y and
z axes (the three panels in Fig.\ \ref{plot4}). We see that shells
with radii $R<60\,$km and $R>250\,$km do not develop any significant
rotational motion during the entire simulation. However,  
at the very early times after core bounce and for radii between 
$\sim$60 km and $\sim$250 km in the shocked, turbulent regions, 
some rotational motion does develop, which with time is 
amplified. By the end of the simulation, those shells have 
acquired spin frequencies that reach $\sim$150 Hz
at a radius of $\sim$100 km, corresponding to a spin period 
of $\sim$7 ms. Furthermore, we notice that although at 
early post-bounce times the sign of the spin vector of those shells kept flipping 
(as indicated by the positive and negative frequency values shown in
Fig.\ \ref{plot4}), at the end of the simulation neighboring shells tended
to align their rotational motion.  The left panel of Fig. \ref{plot5}
depicts the velocity field on a ball at radius 90 km near the end of the simulation.
As this figure clearly shows, coherent fields of rotation have emerged in this 
region (see also the panels in Fig. \ref{plot4}). Near the same final time, 
the shock has been launched into explosion and has 
the multi-turbule morphology portrayed in the right panel of Fig. 
\ref{plot5} as an isoentropy surface.

\begin{figure}
\begin{center}
%\ContinuedFloat
\includegraphics[height=0.6\columnwidth, width=0.75\columnwidth]{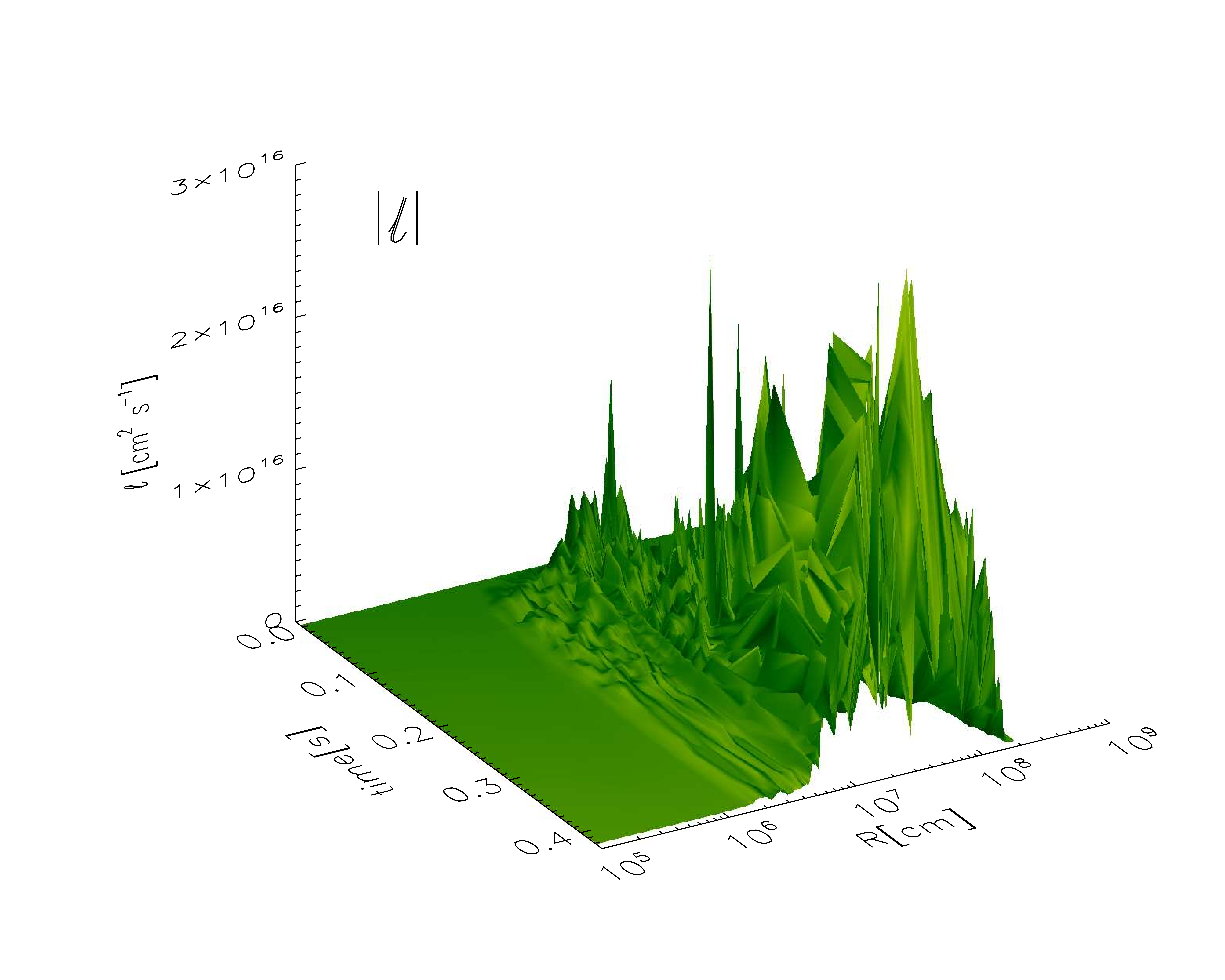}
\includegraphics[height=0.6\columnwidth,
width=0.75\columnwidth]{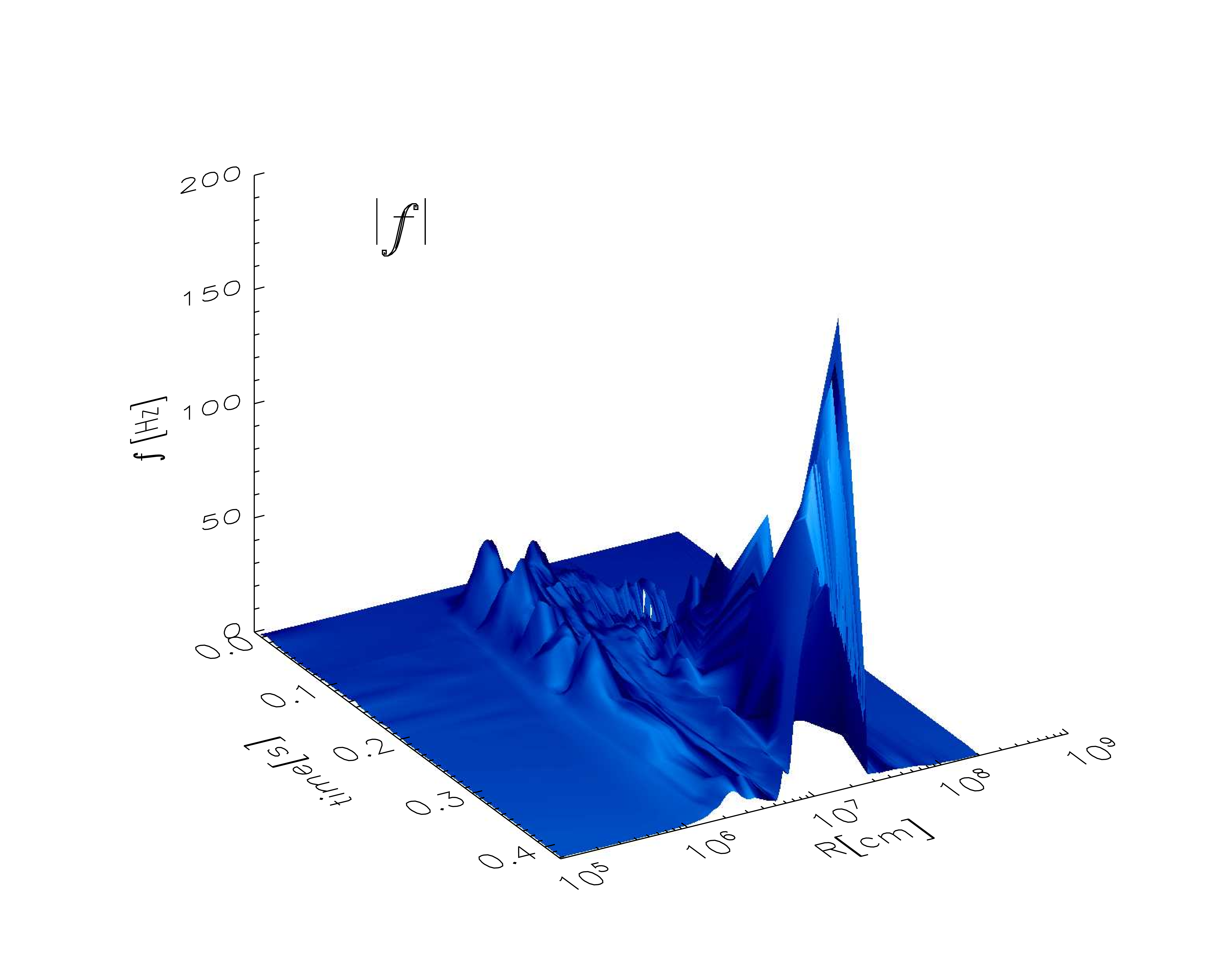}
\end{center}
\caption{Temporal evolution (in seconds) of the magnitude of the specific angular momentum
  (top) and of the magnitude of the rotational frequency (bottom) of spherical 
  shells at various spherical radii R (in centimeters, shown logarithmically). 
  As discussed in the caption to Fig.\ 4 and in the text, individual shells at radii
  $60<R<250\,$km develop significant rotational frequencies over time.
  However, outside this range of radii the rotation 
  induced by non-axisymmetric instabilities is quite modest at all times 
  during the simulation.}
\label{plot6}
\end{figure}

In Fig.\ \ref{plot6} we show the temporal evolution of the magnitude
of the specific angular momentum (top) for various shells at different
radii, and the corresponding magnitude of the rotational frequency (bottom) of those
shells. As in Fig.\ \ref{plot4}, the development of rotational motion
from $\sim$60 to $\sim$250 km is obvious and the high spin rate
of $\sim$150 Hz in this narrow range of radii is manifest.  However,
there is very little mass ($\sim$0.01$-$0.03 M$_{\odot}$) and, hence, total angular momentum 
in this fastest spinning region, and this has major consequences for the 
spin evolution of the entire residual neutron star.

\subsection{Rotational Periods for the Final State}
\label{final_state}

Our simulation shows not only that the inner core isn't spun up 
significantly, but that the mass that does experience an interesting
degree of induced rotation is likely to be ejected with the subsequent blast.
An interesting question arises associated with this ejected angular 
momentum ``bump," shown in Figs.\ \ref{plot1} and \ref{plot2}. If that 
angular momentum were to be accreted later on onto the central regions 
(in a ``fallback" scenario for angular momentum), what would that imply for the terminal 
rotation rate of the nascent neutron star? We address this issue by applying various mass
cuts to the final state of our simulation.  We calculate the
total angular momentum enclosed within a given spherical mass cut (given in column
2 of Table \ref{table1}) and then divide it by a representative moment of inertia 
for the final-state neutron star (taken to be $2 \times 10^{45}\,$ g cm$^{2}$ and 
assuming solid-body rotation) to derive a final angular frequency.
Our results are summarized in Fig.\ \ref{plot7} and Table \ref{table1}. For a range of
mass cuts between $1.2$ and $1.6$ M$_\odot$ (baryonic) for
our ``$1.9\times 10^{52}$ ergs s$^{-1}$" explosion model the final
inferred rotational periods vary between 1.2 and 47 seconds (as can be
seen in Fig.\ \ref{plot7}). The fastest inferred period of $1.2\,$s is
seen for a spherical mass cut of 1.53M$_\odot$ baryonic. Note that 
many measured neutron star masses are near 1.35 M$_{\odot}$ gravitational, 
which translates into $\sim$1.5 M$_{\odot}$ baryonic.  

Apart for the magnitude of such final-state rotational periods, we would also like
to know the direction of rotation and how that varies
with the different spherical mass cuts. We define
two angles to describe the direction of the total angular momentum
enclosed within a certain mass cut. Angle $\theta$
($0^{\circ}\leqslant\theta\leqslant180^{\circ}$) is the polar angle, measured with
respect to the y-axis, and angle $\phi$
($0^{\circ}\leqslant\phi\leqslant360^{\circ}$) is the azimuthal angle, defined
with respect to the z-axis. The two last columns of Table \ref{table1}
show the values of $\theta$ and $\phi$ for the different mass
cuts.  The direction of the angular momentum varies
little for spherical mass cuts from $\sim$1.2 to $\sim$1.4 M$_\odot$).
However, as Table \ref{table1} indicates, it does shows significant 
variation for mass cuts in the range $1.45$ to 1.6 M$_\odot$.

\section{Discussion and Conclusions}
\label{discussion}

\begin{figure}
\begin{center}
%\ContinuedFloat
\includegraphics[height=0.4\columnwidth, width=0.75\columnwidth]{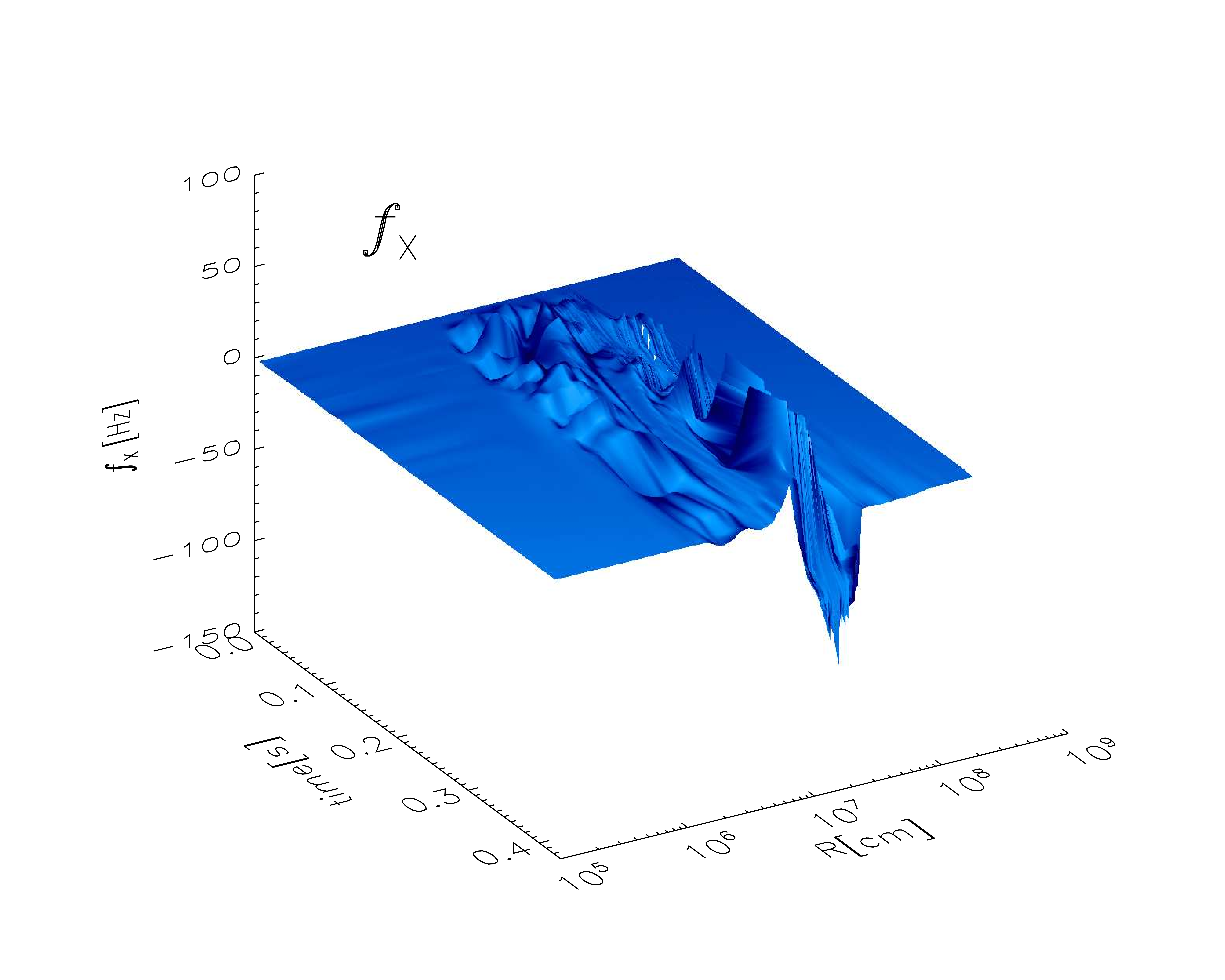}\\
\includegraphics[height=0.4\columnwidth, width=0.75\columnwidth]{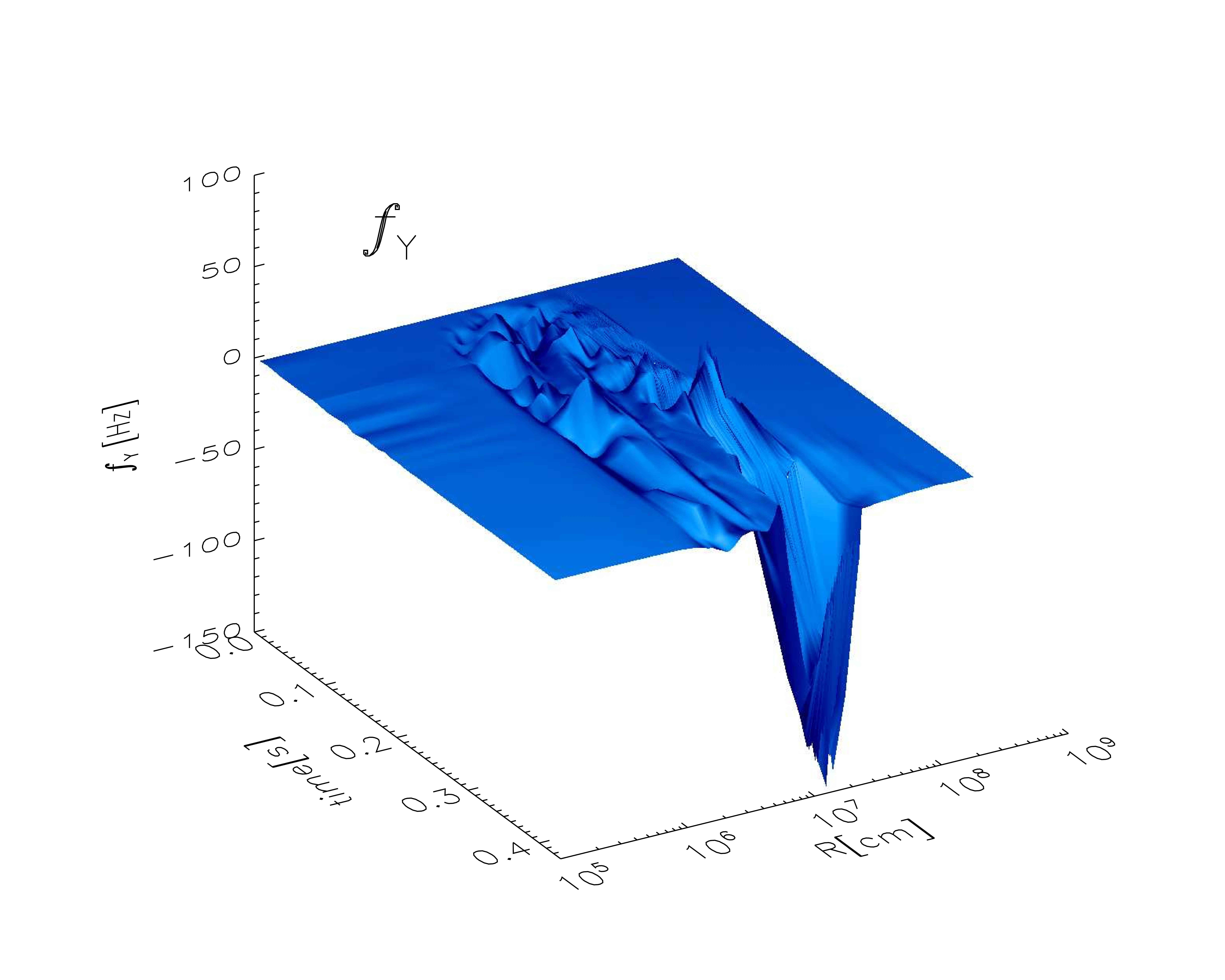}\\
\includegraphics[height=0.4\columnwidth,
width=0.75\columnwidth]{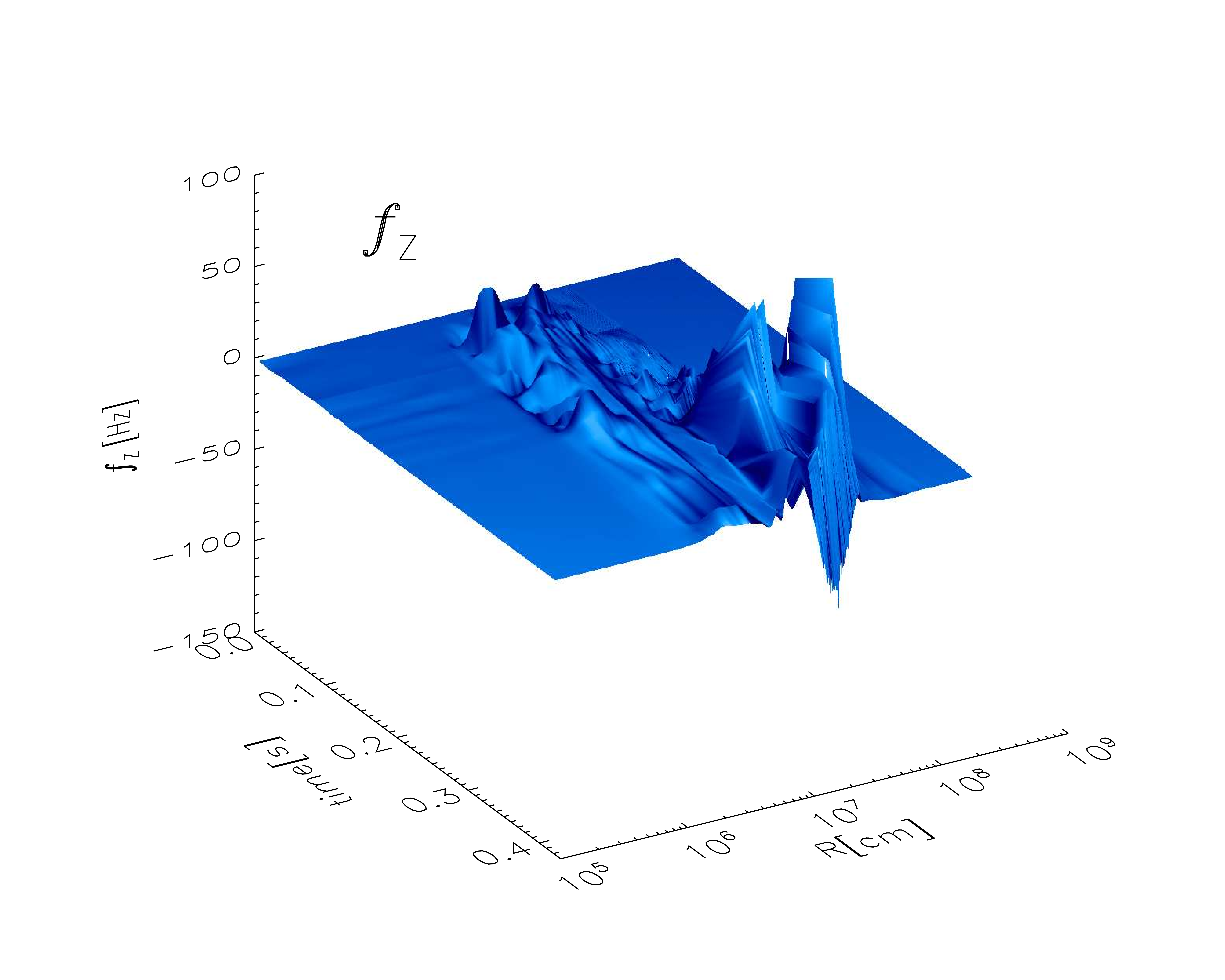}
\end{center}
\caption{The temporal evolution of the spin frequency profiles of thin 
  spherical shells at various spherical radii R (in centimeters, shown
  logarithmically). 
  For each of the individual shells we have calculated the corresponding 
  average rotational frequency, shown here on the vertical axis 
  of each plot. Time is the post-bounce time in seconds. The three panels correspond 
  to rotation around the x, y and z axes (from top to bottom). The 
  positive and negative values of the frequency indicate opposite  
  rotation directions. The computational grid was divided into
  thin spherical shells of width $1\,$km within the inner $250$\ km 
  of the grid and of width $10\,$km for radii between $250$ and $2000\,$km. 
  We notice that for radii in the range  $R<60\,$km and $R>250\,$km, 
  the resulting frequencies remain near ``zero" throughout the entire
  simulation. Interestingly, in the region $60<R<250\,$km, some
  rotational motion develops even during the early stages of the
  simulation and amplifies during the very late stages,
  producing shell rotational frequencies that reach $\sim$$150\,$Hz 
  at a radius of $100\,$km. See text in \S\ref{rot_freq} for a discussion.}
\label{plot4}
\end{figure}

\begin{figure}
\vskip-4in
\begin{center}
\includegraphics[width=0.90\columnwidth]{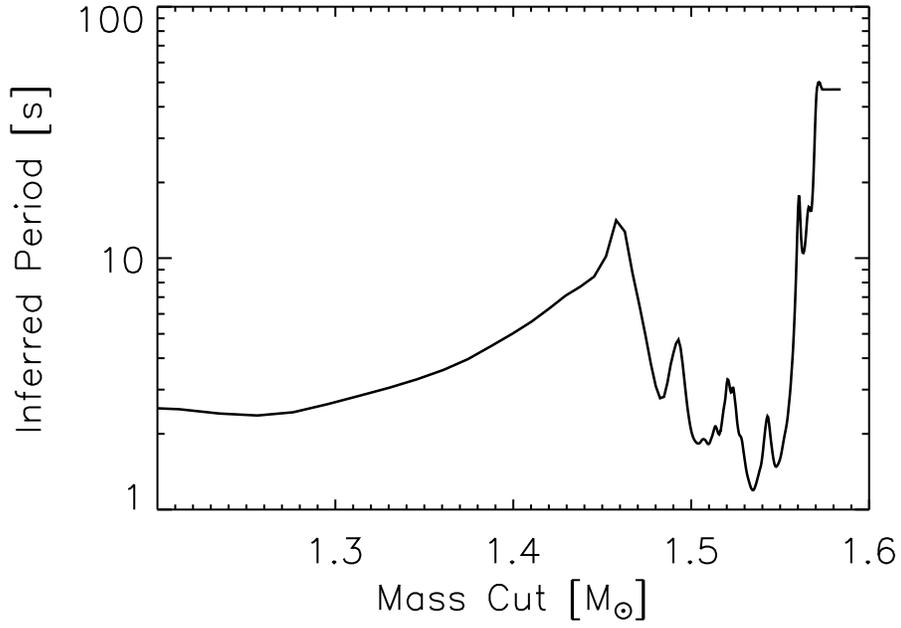} 
\end{center}
\caption{Rotation periods of inferred final-state neutron stars
  as a function of various possible mass cuts (in M$_{\odot}$), 
  as described in the caption to Table 1 and in the text.
  To derive these approximate final spin periods, the enclosed 
  total angular momentum is divided by a final neutron star moment 
  of inertia of $2\times 10^{45}$ g cm$^{2}$. Solid-body rotation 
  for the final state is assumed. The fastest induced spin up is 
  seen to be for a mass cut of $1.53$ M$_\odot$ (baryonic mass), resulting 
  in a slow rotational period of $\sim$1.2 seconds.}
\label{plot7}
\end{figure}

In this paper, we have studied the temporal and spatial evolution of
the rotational profile of a newly-formed neutron star.  The goal was 
to test the possibility that the remnant neutron star created
in the context of core-collapse supernova could be 
induced to rotate rapidly due to the growth of a spiral, non-axisymmetric 
mode of the so-called SASI, even if the progenitor was initially non-rotating.
We employed the 3D hydrodynamics code CASTRO to carry out the simulation, starting 
from the infall phase of a 15-M$_{\odot}$ non-rotating progenitor, 
and followed the core bounce, convection, and explosion stages. 
The core, whose induced rotation was being studied, was included fully 
on the computational grid, a realistic EOS was used, and a simple 
neutrino heating and cooling algorithm enabled us to simulate in 
approximate fashion the turbulent post-bounce delay and subsequent explosive phases.
Our main findings are:

\begin{itemize}
\item  We did not find that a significant spin rate was induced in 
the nascent neutron star. By the end of our simulation the inner core
is seen to rotate very slowly, with spin periods no faster than $\sim$5$-$10 seconds.
 
\item However, in the unstable gain region between the shock and the inner core
(60 km $< R <$ 250 km), counter-rotating shells can acquire enough specific 
angular momentum during the stall phase before explosion that they can achieve
interestingly high spin rates, reaching frequencies around  $\sim$150 Hz (Fig.\ \ref{plot4})
and spin periods between $\sim$6 and $\sim$100 ms. However, the mass in this region,
and the corresponding angular momentum are rather low.

\item At the later stages in the evolution of our simulation, the explosion and 
ejecta transport angular momentum outward behind the exploding shock wave.
Even if some of this  angular momentum were to fall back at later times, the 
minimum spin period that the residual neutron star could achieve 
is likely to be quite modest (Fig.\ \ref{plot7} and Table \ref{table1}).

\end{itemize}

Blondin \& Mezzacappa (2007) employed a gamma-law equation of state,
did not follow the hydrodynamics of the inner core, did not include in their
simulations the collapse or the explosion phases, and did not include
any neutrino interactions.  Yet, they concluded that even a non-rotating
progenitor could in principle leave behind a rotating remnant. In this
qualitative conclusion we agree with Blondin \& Mezzacappa (2007).

However, we used a general nuclear equation of state, included the entire
core all the way to the center in the hydrodynamic simulation, included in
a crude fashion the effects of neutrino heating, cooling, and electron
capture, and followed both the collapse and the explosion phases. What we
found was that the degree of induced central rotation of the residual
proto-neutron star, after explosion and for any given degree of fallback,
produces periods between $\sim$1 and $\sim$10 seconds. Nevertheless, those authors
suggest that a longer delay could lead to even greater net spin rates and
see in their simulation a significant ramp up from $\sim$0.4 to $\sim$0.9
seconds after bounce. However, they concede that the spiral mode spin-up
phenomenon should not continue beyond the onset of explosion. Since our
explosion occurs near $\sim$250 ms, one could argue that a longer delay
could result in a larger induced spin.  While this seems plausible,
\cite{wongwa} have performed a new set of approximate 3D simulations of
core-collapse supernovae with non-rotating progenitors, focusing on the
remnant neutron star kick velocities (see also Nordhaus et al. 2010b).
They also show that after as much as $\sim$1.4 seconds after bounce the
induced spin rates are still fairly low, with final periods in the range
$\sim$500 to $\sim$1000 ms. Such long periods are approximately in accord
with those we derive here in our ``optimum" case.

In summary, the increase in the spin rate from the optimum derived 
from our simulation, which terminated at $\sim$0.422 seconds,
to that seen in the simulations of \cite{wongwa}, which terminated at $\sim$1.4 seconds,
was a factor of $\sim$2. The calculations performed by Blondin \& Mezzacappa (2007)
implied that the enhancement in delaying from $\sim$0.422 seconds to
$\sim$0.9 seconds might be a factor of $\sim$10 or more.
We take this as a further indication that induced rotation by spiral 
modes may not be adequate to explain the observed (or inferred) 
rotation rates of most pulsars. Nevertheless, induced spin remains 
an intriguing, if sub-dominant, potential contribution to the overall
angular momentum budget of nascent neutron stars.  Where final 
neutron-star spin periods of seconds are observed, the spiral-mode 
spin-up mechanism may remain viable.  

One could argue that the delay to explosion we 
witness is already rather long. We suggest that in order for the 
neutrino mechanism to be robust and to generate explosion energies 
sufficient to overcome the binding energies of the progenitor mantle, while 
at the same time yielding explosion energies at infinity of $\sim$10$^{51}$ ergs,
the onset of explosion should probably be rather early after bounce.  However, 
this has yet to be demonstrated. We emphasize that our models start with zero 
angular momentum. A longer delay and total angular momentum conservation would naturally make it less
and less likely that any significant {\it net} angular momentum would be left behind.
This effect is implicit in Fig. \ref{plot7}, for which the residual net spin period
jumps up on the right-hand side at greater mass cuts. 
Individual zones could initially be left with interesting spins, but others would have to 
be in the opposite directions required by net angular momentum conservation.
At very late times, the residue would have to be in solid-body rotation at low net 
angular momentum.

In this paper, we have not explored various progenitor masses,
nor a range of initial spin structures. Our goal was to determine whether
a sizable rotation rate could be induced in the residual proto-neutron
star when the progenitor itself was non-rotating. The behavior
of the modest spiral mode we do see, and the overall role of rotation
in the supernova phenomenon itself, remain to be fully mapped out.
As computational capabilities improve, and codes acquire more physics
and sophistication, all these issues can and will be readdressed.

\acknowledgments

The authors would like to acknowledge Rodrigo
Fern\`andez, Timothy Brandt, John Bell, and Brian Metzger for fruitful discussions and input.
A.B. and J.N. are supported by the Scientific Discovery through
Advanced Computing (SciDAC) program of the DOE, under grant number
DE-FG02-08ER41544.  E.R. is supported by the NSF under the subaward no. ND201387 to the
Joint Institute for Nuclear Astrophysics (JINA, NSF PHY-0822648), and
A.B. receives support from the NSF PetaApps program, under award OCI-0905046 via a subaward
no. 44592 from Louisiana State University to Princeton University. 
Work at LBNL was supported in part by the SciDAC Program under contract DE-AC02-05CH11231.
The authors would like to thank the members of the Center for
Computational Sciences and Engineering (CCSE) at LBNL for their
invaluable support for CASTRO. The authors employed computational
resources provided by the TIGRESS high performance computer center at
Princeton University, which is jointly supported by the Princeton
Institute for Computational Science and Engineering (PICSciE) and the
Princeton University Office of Information Technology; by the National
Energy Research Scientific Computing Center (NERSC), which is
supported by the Office of Science of the US Department of Energy
under contract DE-AC03-76SF00098 and on the Kraken and Ranger supercomputers, 
hosted at NICS and TACC and provided by the National Science Foundation through
the TeraGrid Advanced Support Program under grant number TGAST100001.

\begin{deluxetable}{cccccc}

\tablewidth{500pt}

\tablecaption{Results at the Final Stage Of The Simulation\tablenotemark{(a)}}
\tablehead{Spherical Mass Cut (Baryonic) &$L$ &  P & $f$  & $\theta$ &$\phi$
\\ 
$[M_\odot]$ & $[\times10^{46}$g cm$^2$s$^{-1}]$ & $[$s$]$ & $[$Hz$]$
& $[^o]$ & $[^o]$}
\startdata
1.20   & 0.501    & 2.50 &0.40 &
97.6 & 67.2 \\
 1.25& 0.531   & 2.37
&0.42  & 92.1 & 67.5\\
1.30  & 0.479   & 2.83
&0.35  & 89.7& 70.5\\
1.35  & 0.380   & 3.30
&0.30  & 96.7& 84.2\\
1.40  & 0.252  & 4.99
&0.20  &111.6 & 272.3\\
1.45   & 0.125   & 10.17
&0.1  & 152.5&0.25\\
1.50   & 0.619   & 2.02
&0.49  & 128.6& 51.6\\
1.55   & 0.790   & 1.59
&0.63  & 21.6 & 5.3\\
1.60  & 0.027   & 46.89
&0.02  &125.5 & 191.9
\enddata
\label{table1}
\tablenotetext{(a)}{At $t=0.422\,$s after core bounce; In column 1, we 
  list various spherical mass cuts. The total angular momentum
  enclosed within that mass is given in column 2. If this enclosed
  angular momentum were to be left with the nascent neutron star,
  which we assume to have a moment of inertia of $I\sim2\times10^{45}$g cm$^2$, then the resulting rotational
periods and frequencies would be those presented in columns 3 and 4,
respectively. The direction of the total enclosed angular momentum $L$ for each
case is represented by two angles given in the last two columns. The
polar angle $\theta$ ($0^o\leqslant\theta\leqslant180^o$) is defined with
respect to the y-axis and the azimuthal angle $\phi$
($0^o\leqslant\phi\leqslant360^o$) is measured with respect to the
z-direction.}
\end{deluxetable}

\clearpage

\thispagestyle{empty}

% figure 1

\clearpage

% figure 2

\clearpage

% figure 3

\clearpage

% figure 4

\clearpage

% figure 5

\clearpage

% figure 6

%\clearpage

% figure 7


\begin{thebibliography}{}
\bibitem[Almgren et al.(2010)]{almgren} Almgren, A.S., Beckner, V.E.,
Bell, J.B., Day, M.S., Howell, L.H., Joggerst, C.C., Lijewski, M.J., Nonaka, A.,
Singer, M., \& Zingale, M. 2010, \apj, 715, 1221
\bibitem[Atoyan (1999)]{atoyan}
Atoyan, B. M.\ 1999, \aap, 346, L49
%\bibitem[Arzoumanian et al.\ (2002)]{arzoumanian}
%Arzoumanian, Z., Chernoff, D. F., \& Cordes, J. M.\ 2002, \apj, 568,
%289
%\bibitem[Bruenn et al.\ (2010)]{bruenn10}
%Bruenn, S. W., Mezzacappa, A., Hix, W. R., Blondin, J. M., Marronetti,
%P., Messer, O. E. B., Dirk, C. J., \& Yoshida, S.\ 2010, arXiv:1002.4914
%\bibitem[Burrows et al.\ (2006)]{burrows06}
%Burrows, A., Livne, E., Dessart, L., Ott, C. D. \& Murphy, J. \ 2006,
%\apj, 640, 878
%\bibitem[Burrows et al.\ (2007)]{burrows07}
%|. 2007, \apj, 655, 416
\bibitem[Blondin \& Mezzacappa (2007)]{BM07} 
Blondin, J. M., \& Mezzacappa, A.\ 2007, Nature, 445, 58
\bibitem[Blondin \& Shaw (2007)]{BS07}
Blondin, J. M., \& Shaw, S.\ 2007, \apj, 656, 366
%\bibitem[Camilo et al.\ (1993)]{camilo93}
%Camilo, F., Nice, D.J., \& Taylor, J.H.\ 1993, \apjl, 412, L37
%\bibitem[Chakrabarty (2008)]{chakra}
%Chakrabarty, D.\ 2008, AIP Conf. Proc., 1068, 67
\bibitem[Chevalier \& Emmering(1986)]{chevalier} Chevalier, R. \& Emmering, R.T. 1986, \apj, 304, 140
\bibitem[Fern\`andez (2010)]{fernandez}
Fern\`andez, R. \ 2010, arXiv:1003.1730
%\bibitem[Heger et al.\ (2004)]{heger04}
%Heger, A., Woosley, S. E., Langer, N., \& Spruit, H. C.\ 2004, IAUS, 215,
%519
\bibitem[Heger et al.\ (2005)]{heger05}
Heger, A., Woosley, S. E., \& Spruit, H. C.\ 2005, \apj, 626, 350
%\bibitem[Hessels (2009)]{hessels}
%Hessels, J. W. T. \ 2009, arXiv:0903.0493
\bibitem[Hirschi et al.\ (2004)]{hirschi}
Hirschi, R., Meynet, G., \& Maeder, A.\ 2004, \aap, 425, 649
%\bibitem[Janka et al.\ (2004)]{janka04}
%Janka, H.-Th., Scheck, L., Kifonidis, K., M\"uller, E., \& Plewa,T. \
%2004, ASPC, 332, 363J  
%\bibitem[Janka et al.\ (2007)]{janka07}
%Janka, H.-Th., Langanke, K., Marek, A., Martínez-Pinedo, G.,\&
%M\"uller, B. \ 2007, Phys. Repts., 442, 38
%\bibitem[Kaplan et al.\ (2008)]{kaplan}
%Kaplan, D. L., Chaterjee, S., Gaensler, B. M. \& Anderson, L.\ 2008, \apj, 677, 1201
%\bibitem[Kiziltan \& Thorsett (2009)]{kiziltan}
%Kiziltan, B. \& Thorsett, S. E.\ 2009, \apj, 693L, 109
%\bibitem[Kluzniak et al.\ (1988)]{kluzniak88}
%Kluzniak, W., Ruderman, M., Shaham, J., \& Tavani, M.\ 1988, Nature,
%334, 225
%\bibitem[Lamb \& Yu (2005)]{lamb05}
%Lamb, F. K. \& Yu, W.\ 2005, ASP Conference Series, 328
\bibitem[Liebend\"orfer(2005)]{lieben} Liebend\"orfer, M. 2005, \apj, 633, 1042
%\bibitem[Lorimer (2008)]{lorimer08}
%Lorimer, D. R.\ 2008, LRR, 11, 8
\bibitem[Lorimer (2009)]{lorimer09}
Lorimer, D. R.\ 2009, Astrophys. \& Space Sci. Lib., 357, 1
\bibitem[Lorimer (2010)]{lorimer10}
Lorimer, D. R.\ 2010, to appear in the Proceedings of the ICREA Workshop 
on The High-Energy Emission from Pulsars and their Systems, 
held in Sant Cugat, Spain, 2010 April 12-16 (Springer), arXiv:1008.1928 
\bibitem[Maeder \& Meynet (2000)]{mm00}
Maeder, A. \& Meynet, G.\ 2000, ARA\&A, 38, 143
\bibitem[Maeder \& Meynet (2004)]{mm04}
Maeder, A. \& Meynet, G.\ 2004, \aap, 422, 225  %  IAUS, 215, 500
\bibitem[Murphy \& Burrows (2008)]{murphy08}
Murphy, J. W., \& Burrows, A.\ 2008, ApJ, 688, 1159
\bibitem[Narayan(1987)]{narayan} Narayan, R. 1987, \apj, 319, 162
%\bibitem[Ng \& Romani (2007)]{ng}
%Ng, C.-Y. \& Romani, R.\ 2007, \apj, 660, 1357
\bibitem[Nordhaus et al.\ (2010a)]{jason}
Nordhaus, J., Burrows, A., Almgren, A., \& Bell, J.\ 2010a, \apj, 720, 694
\bibitem[Nordhaus et al.\ (2010b)]{jason2}
Nordhaus, J., Brandt, T., Burrows, A., Livne, E. \& Ott, C.D. 2010b, accepted to \prd, arXiv:1010.0674
\bibitem[Ott et al.\ (2006a)]{ott06a}
Ott, C. D., Burrows, A., Thompson, T. A., Livne, E., \& Walder, R.\
2006a, \apjs, 164, 130
\bibitem[Ott et al.\ (2006b)]{ott06b}
Ott, C. D., Burrows, A., Livne, E., \& Walder, R.\ 2006b, \apj, 600,
834
%\bibitem[Ott et al.\ (2008)]{ott08}
%Ott, C.D., Burrows, A., Dessart, L., Livne, E.\ 2008, \apj, 685, 1069
%\bibitem[Perna et al.\ (2008)]{perna}
%Perna, R., Soria, R., Pooley, D., \& Stella, L. \ 2008, MNRAS, 384, 1638
%\bibitem[Scheck et al.\ (2004)]{scheck}
%Scheck, L., Plewa, T., Janka, H.-Th., Kifonidis, K. \& M\"uller,
%E. \ 2004, PhRvL, 92, 1103 
\bibitem[Shen et al.\ (1998a)]{shena}
Shen, H., Toki, H., Oyamatsu, K., \& Sumiyoshi, K.\ 1998a, Nucl. Phys. A, 637, 435
\bibitem[Shen et al.\ (1998b)]{shenb}
|. 1998b, Prog. Th. Phys., 100, 1013
%\bibitem[Sigurdsson \& Phinney (1995)]{sigurdsson95}
%Sigurdsson, S., \& Phinney, E.S.\ 1995, \apjs, 99, 609
%\bibitem[Spruit \& Phinney (1998)]{spruit}
%Spruit, H. \& Phinney, E. S.\ 1998, Nature, 393, 139
%\bibitem[Whitehouse \& Liebend\"orfer (2008)]{whitehouse}
%Whitehouse, S. \&  Liebend\"orfer, M.\ 2008, nuco.confE.243W
\bibitem[Wongwathanarat et al.\ (2010)]{wongwa}
Wongwathanarat, A., Janka, H.-Th. \& M\"uller, E. \ 2010, submitted to \apj, arXiv:1010.0167
\bibitem[Woosley \& Weaver (1995)]{woos95}
Woosley, S. E. \& Weaver, T. A.\ 1995, \apjs, 101, 181
\end{thebibliography}
\end{document}